\documentclass[aps,prl,twocolumn,superscriptaddress,noeprint]{revtex4-1}
\usepackage{graphicx}
\usepackage{physics}
\usepackage{multirow}
\begin{document}

\title{Chiral transport along magnetic domain walls in the quantum anomalous Hall effect}

\author{I.\ T.\ Rosen}
\affiliation{Department of Applied Physics, Stanford University, Stanford, California 94305, USA}
\affiliation{Stanford Institute for Materials and Energy Sciences, SLAC National Accelerator Laboratory, Menlo Park, California 94025, USA}
\author{E.\ J.\ Fox}
\affiliation{Department of Physics, Stanford University, Stanford, California 94305, USA}
\affiliation{Stanford Institute for Materials and Energy Sciences, SLAC National Accelerator Laboratory, Menlo Park, California 94025, USA}
\author{X.\ Kou}
\affiliation{Department of Electrical Engineering, University of California, Los Angeles 90095, USA}
\affiliation{School of Information Science and Technology, ShanghaiTech University 201210, China}
\author{L.\ Pan}
\affiliation{Department of Electrical Engineering, University of California, Los Angeles 90095, USA}
\author{K.\ L.\ Wang}
\affiliation{Department of Electrical Engineering, University of California, Los Angeles 90095, USA}
\author{D.\ Goldhaber-Gordon}
\email[E-mail: ]{goldhaber-gordon@stanford.edu}
\affiliation{Department of Physics, Stanford University, Stanford, California 94305, USA}
\affiliation{Stanford Institute for Materials and Energy Sciences, SLAC National Accelerator Laboratory, Menlo Park, California 94025, USA}

\maketitle

\textbf{
The recent prediction~\cite{Yu2010}, and subsequent discovery~\cite{Chang2013}, of the quantum anomalous Hall (QAH) effect in thin films of the three-dimensional ferromagnetic topological insulator (MTI) (Cr$_y$Bi$_x$Sb$_{1-x-y}$)$_2$Te$_3$ has opened new possibilities for chiral-edge-state-based devices in zero external magnetic field. Like the $\nu=1$ quantum Hall (QH) system, the QAH system is predicted to have a single chiral edge mode circulating along the boundary of the film. Backscattering of the chiral edge mode should be suppressed, as recently verified by the observation of well-quantized Hall resistivities $\rho_{yx}=\pm h/e^2$, along with longitudinal resistivities as low as a few ohms~\cite{Bestwick2015,Chang2015}. Dissipationless 1D conduction is also expected along magnetic domain walls~\cite{Qi2009,Wang2014,Checkelsky2012,Wakatsuki2015}. Here, we intentionally create a magnetic domain wall in a MTI and study electrical transport along the domain wall. We present the first observation of chiral transport along domain walls, in agreement with theoretical predictions. We present further evidence that two modes equilibrate and co-propagate along the length of the domain wall.
}

Edge conduction in the QAH system is a consequence of the system's topological non-triviality~\cite{Qi2006}. The QAH system is topologically classified by the Chern number $C=\pm1$, corresponding to upwards or downwards magnetization, respectively. At the interface between MTI and vacuum, the Chern number transitions from $C=\pm1$ to the topologically trivial $C=0$; one chiral edge mode propagates along this interface. At a magnetic domain wall, however, the Chern number changes by two, from $C=+1$ to $C=-1$. Accordingly, two chiral modes should co-propagate along this interface~\cite{Nomura2011,Upadhyaya2016,Wang2014}. If the domain does not reach the film's edge, the modes at the domain wall should simply circle the domain, having no effect on transport. But if the domain wall connects two of the device's edges or contacts, it may affect two- and/or four-terminal resistances. Analogous transport has been studied in the QH effect in graphene-based two-dimensional electron gases, where patterned gates can produce adjacent regions of different density and hence different filling factor~\cite{Williams2007,Amet2014}.

Previously, the magnetization of MTIs has been flipped without spatial control, by sweeping a homogenous external magnetic field through the coercive field $H_C^{\textrm{TI}}$. For most MTIs that display the QAH effect, $\rho_{yx}$ transitions from $\mp h/e^2$ to $\pm h/e^2$ over a substantial range of field ($H=150$ to $H=200$ mT for the material used in the work), and $\rho_{xx}$ has a maximum in this field range~\cite{Checkelsky2014,Kou2014,Mogi2015,Kandala2015,Lachman2015,Kou2015a}. Hysteresis loops of the four-terminal resistances of a 50 $\mu$m wide Hall bar of MTI film are shown in Fig.~1a. In some samples and in some temperature ranges, $R_{yx}$ jumps in discrete steps when the external field is swept through $H_C^{\textrm{TI}}$~\cite{Liu2016a,supinfo}. Each jump likely represents rearrangement of the magnetic domain structure. Since within a domain the bulk of a MTI is highly insulating at the lowest temperatures~\cite{Bestwick2015,supinfo}, these jumps suggest that domain walls host conductive modes. The set of discrete $R_{yx}$ values is not reproducible between separate magnetic field sweeps, suggesting that the network of magnetic domain walls is complex.

To better study transport along domain walls, we intentionally engineered a magnetic domain wall in 6 quintuple layer (Cr$_{0.12}$Bi$_{0.26}$Sb$_{0.62}$)$_2$Te$_3$ by spatially modulating the external magnetic field $H$, and then measured electronic transport along the domain wall at $H=0$. Our results confirm chiral transport along magnetic domain walls in the quantum anomalous Hall system, and support the prediction~\cite{Upadhyaya2016} that domain walls host two co-propagating modes whose carriers fully equilibrate.

We spatially modulated the magnetic field applied to the MTI film using Meissner repulsion from a bulk superconductor~\cite{Carmona1995}. A niobium cylinder of 1.5 mm diameter and 2 mm height was placed partially covering the surface of a patterned MTI film, but not in electrical contact. When superconducting, the niobium cylinder screens a portion of the external magnetic field, as sketched in Fig.~1b. Far from the cylinder, the external magnetic field is unscreened and the magnetization of the MTI film switches direction when the external field reaches $H^{\textrm{TI}}_C$. The screened portion of the MTI film, underneath the cylinder, does not switch magnetization until the external magnetic field reaches a higher value. We define $M_z=M/\abs{M}=1\,(-1)$ to represent upwards- (downwards-) pointing magnetization of the MTI film far from the superconductor. $M_z^{\textrm{SC}}$ represents the equivalent quantity in the screened region of the MTI.

\begin{figure}
\centering
\includegraphics[width=0.45\textwidth]{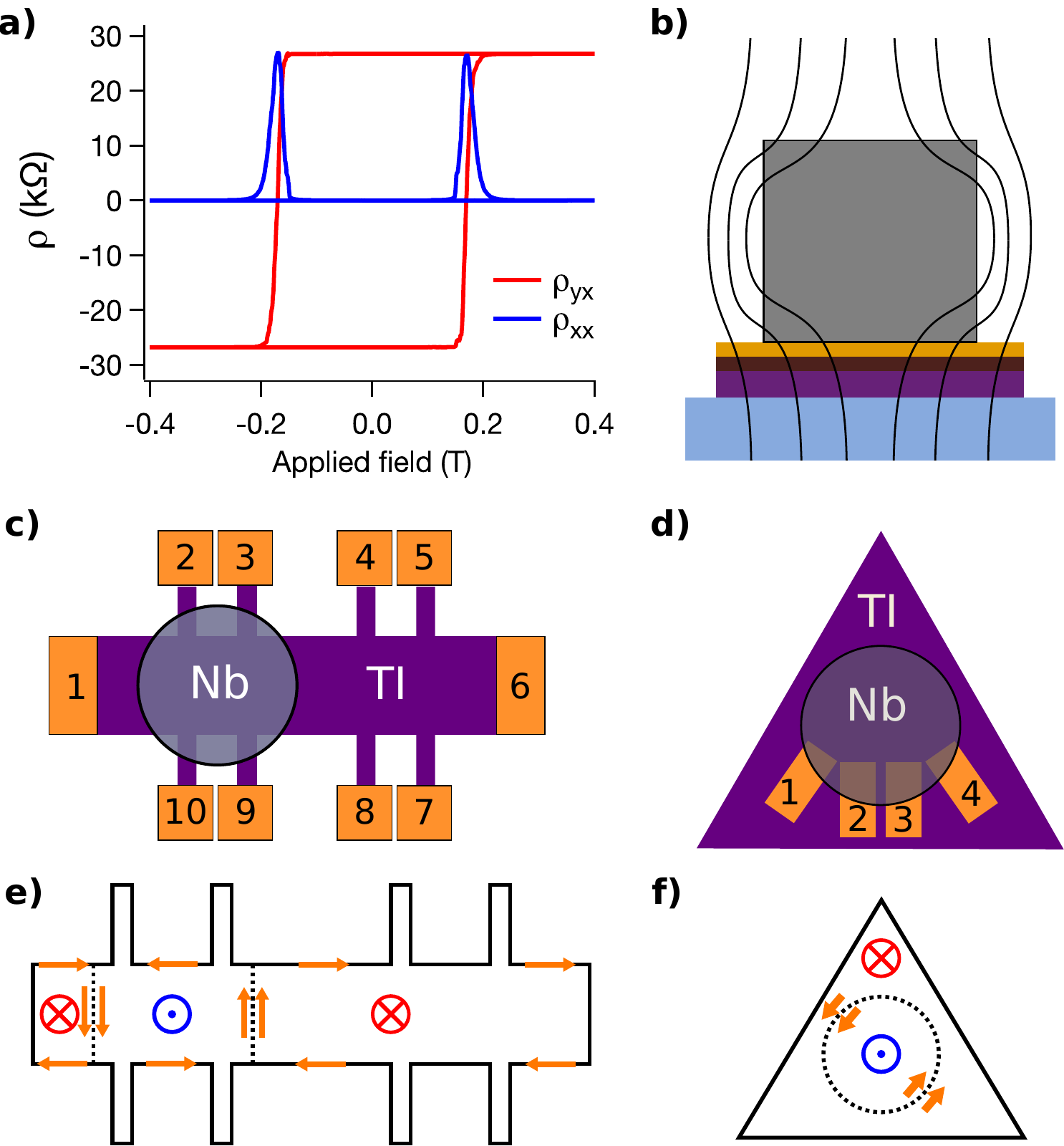}
\label{fig1}
\caption{a) Hysteresis curves of the Hall (red) and longitudinal (blue) resistivities in a top gated Hall bar of the magnetic topological insulator (Cr$_{0.12}$Bi$_{0.26}$Sb$_{0.62}$)$_2$Te$_3$. b) Schematic side view of a superconducting cylinder screening magnetic flux lines away from a thin MTI film (purple) on a GaAs substrate (blue). The cylinder and MTI are separated by an alumina dielectric (brown) and a top gate (gold). Thicknesses are not to scale. c) Schematic top view of Device A. d) Schematic top view of Device B. e), f) Schematics of the expected chiral electrical transport in the presence of magnetic domain walls (dotted lines), when $M_z=-1$ and $M_z^{\textrm{SC}}=1$, in Devices A and B. The orange arrows indicate a single chiral mode each, and the red and blue vectors indicate the direction of magnetization.}
\end{figure}

To create a domain wall, the MTI film is first fully magnetized downwards $M_z=M_z^{\textrm{SC}}=-1$ with a large external field. A positive field slightly above $H_C^{\textrm{TI}}$, typically between $\mu_0 H=180$ mT and 200 mT, is then applied. Only the unscreened region switches magnetization, so that $M_z=1$, and $M_z^{\textrm{SC}}=-1$. This process forms a domain wall near the boundary of the superconductor; its exact position depends on the geometric demagnetization of the superconductor. Applying oppositely signed fields forms the magnetic configuration $M_z=-1$, $M_z^{\textrm{SC}}=1$. Domains were formed with the sample held between 25 and 200 mK.

The transport properties of the domain wall were studied at zero field in two geometries. Device A (Fig.~1c, schematic) is a Hall bar with eight voltage terminals and two current contacts. A niobium cylinder was placed on the device's surface, covering the leftmost four voltage terminals. Device B (Fig.~1d, schematic) consists of four contacts inside a large region of MTI. Since the contacts are not connected by an edge of the MTI, when Device B was uniformly magnetized, no conductive channel connected the contacts, and the resistance between pairs of contacts exceeded $8\ \textrm{M}\Omega$ at $25$ mK. The boundary of a niobium cylinder overlaps all four contacts to form a magnetic domain wall connecting the contacts. The MTI's Fermi level in Device B was tuned with an electrostatic top gate; Device A's top gate, however, was unintentionally shorted to one contact and was left at zero volts during measurements.

Device A was first fully magnetized downwards by an external field $\mu_0 H=-1$ T and then returned to $\mu_0 H=0$. At base temperature, the Hall resistance approached $-h/e^2$ both underneath and away from the superconducting cylinder, as indicated in Table~I. In both regions, the longitudinal resistance was small compared to $h/e^2$, indicating nearly dissipationless conduction along the edges of the device.

\begin{table}[h]
\centering
\begin{tabular}{*{2}{c} *{2}{l} c c}
$M_z$ & $M_z^{SC}$ & Measurement & Resistance & Predicted & Measured \\
\hline
\multirow{4}{*}
{$-$1} & \multirow{4}{*} {$-$1} & $R_{xx}$ & $R_{16,45}$ & 0 & 0.005\\
 & & $R_{xx}^{\textrm{SC}}$ & $R_{16,23}$ & 0 & 0.004\\
 & & $R_{yx}$ & $R_{16,84}$ & $-$1 & $-$0.999\\
 & & $R_{yx}^{\textrm{SC}}$ & $R_{16,93}$ & $-$1 & $-$1.002\\
\hline
\multirow{6}{*}
{$-$1} & \multirow{6}{*}
{1} & $R_{xx}$ & $R_{16,45}$ & 0 & 0.215\\
 & & $R_{xx}^{\textrm{SC}}$ & $R_{16,23}$ & 0 & 0.018\\
 & & $R_{yx}$ & $R_{16,84}$ & $-$1 & $-$0.989\\
 & & $R_{yx}^{\textrm{SC}}$ & $R_{16,93}$ & 1 & 1.006\\
 & & $R_{xx}^{\textrm{top}}$ & $R_{16,34}$ & 0 & 0.207\\
 & & $R_{xx}^{\textrm{bottom}}$ & $R_{16,98}$ & 2 & 2.149\\
\hline
\multirow{6}{*}
{1} & \multirow{6}{*}
{$-$1} & $R_{xx}$ & $R_{16,45}$ & 0 & 0.015\\
 & & $R_{xx}^{SC}$ & $R_{16,23}$ & 0 & 0.006\\
 & & $R_{yx}$ & $R_{16,84}$ & 1 & 0.997\\
 & & $R_{yx}^{SC}$ & $R_{16,93}$ & $-$1 & $-$0.998\\
 & & $R_{xx}^{\textrm{top}}$ & $R_{16,34}$ & 2 & 2.348\\
 & & $R_{xx}^{\textrm{bottom}}$ & $R_{16,98}$ & 0 & 0.340\\
\end{tabular}\\
\caption{Four-terminal resistances in Device A, measured at zero field and 30 mK. The resistance $R_{ij,kl}$ indicates the four-terminal resistance given by sourcing current between terminals $i$ and $j$, and measuring the voltage between terminals $k$ and $l$. The magnetic configuration is indicated by $M_z$ ($M_z^{\textrm{SC}}$), the sign of out-of-plane magnetization outside (inside) of the region screened by the superconducting cylinder. Resistance values are given in units of $h/e^2$. Predicted values, given by the Landauer-B\"uttiker formalism, assume full equilibration of carriers propagating along domain walls\cite{supinfo}.}
\label{tablea}
\end{table}

Next, the external magnetic field was swept, as detailed previously, to attempt to create a magnetic domain. After returning to zero field, the Hall resistivity in the screened and unscreened regions of Device A had opposite signs, confirming the formation of distinct magnetic domains in the Hall bar. The longitudinal resistance was small compared to $h/e^2$ inside both regions, indicating nearly dissipationless conduction along edges within the domains. Were the Fermi level optimized by the top gate, the longitudinal resistance might have been further reduced. Four-terminal resistances in Device A, in various magnetic configurations, are presented in Table~\ref{tablea}.

Having confirmed the creation of a magnetic domain wall, we study the equilibration of the two chiral modes expected to co-propagate along the domain wall. Consider carriers traveling along the domain walls sketched in Fig.~1e. For sufficiently long domain walls, we expect full equilibration, meaning that carriers leaving the domain walls move rightwards and leftwards along the device's edges with equal probability~\cite{Upadhyaya2016}. For full equilibration, the Landauer-B\"uttiker formalism predicts four-terminal longitudinal resistances $R^{\textrm{top}}_{xx}$ and $R^{\textrm{bottom}}_{xx}$, measured \emph{across} the domain wall, of $0$ and $2h/e^2$~\cite{supinfo}, in reasonable agreement with our results as shown in Table~\ref{tablea}. The measured resistances slightly exceed the predicted values likely because $R_{xx}$ is not exactly zero within each domain, though slightly imperfect equilibration would also have this effect~\cite{supinfo}.

Using Device B, we examine how carriers propagate along domain walls. The effective longitudinal resistance $R_{14,23}$ quantifies dissipation along the domain wall, and the effective Hall resistance $R_{13,24}$ and its mirror $R_{42,31}$ establish the chirality of the conductive modes. These four-terminal measurements are not standard longitudinal and Hall measurements, but are topologically analogous to typical $R_{xx}$ and $R_{yx}$ measurements in the $\nu=2$ QH system if indeed two chiral modes co-propagate along the interface between two insulating domains. $R_{14,23}$, shown in Fig.~2a as a function of gate voltage, was smallest when the gate voltage was $-8$ V; here, the Fermi level presumably sits in the center of the gap. The low longitudinal resistance at the optimum gate voltage, as shown in Table~II, indicates nearly dissipationless conduction along the domain wall.

\begin{figure*}
\centering
\includegraphics[width=0.85\textwidth]{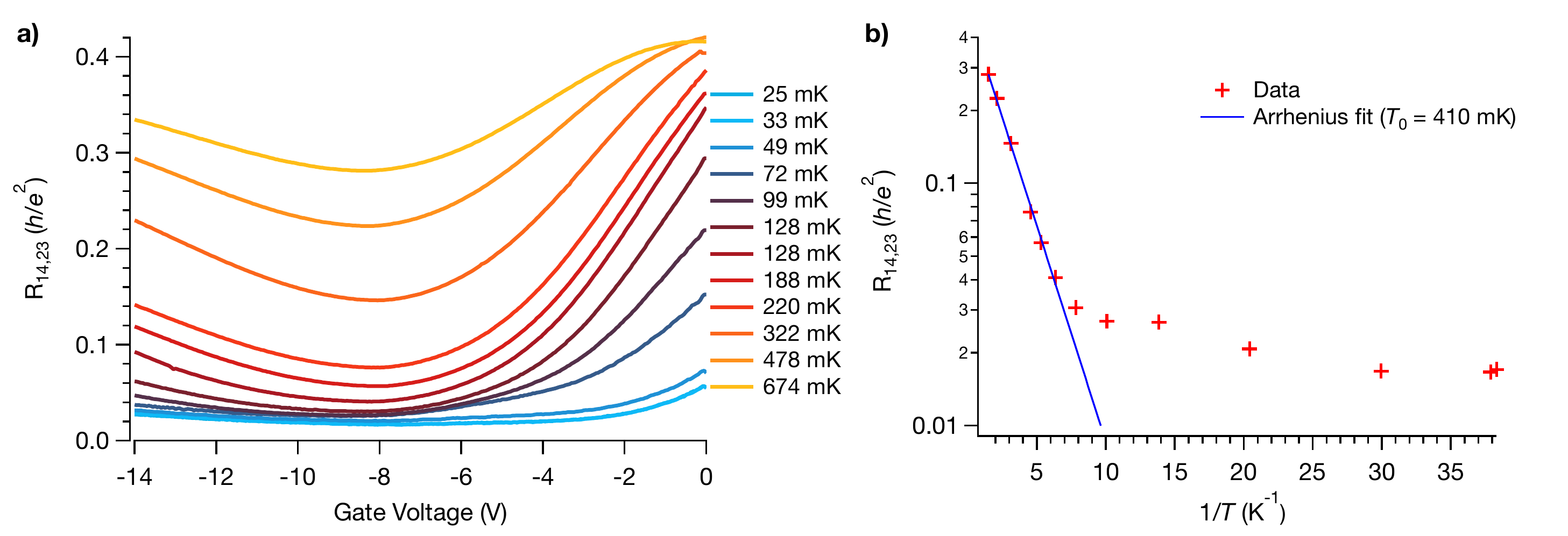}
\label{fig2}
\caption{a) Longitudinal resistance $R_{14,23}$ of Device B as a function of gate voltage at temperatures between 25 mK and 675 mK. Magnetization was $M_z=1$ and $M_z^{\textrm{SC}}=-1$, such that a domain wall with clockwise chirality connects the contacts. b) Longitudinal resistance $R_{14,23}$ of Device B at gate voltage $-8$ V, shown on an Arrhenius plot. Only data at temperatures exceeding 150 mK are included in the fit to an Arrhenius law.}
\end{figure*}

\begin{table}[h]
\centering
\begin{tabular}{*{2}{c} *{2}{l} c c}
$M_z$ & $M_z^{SC}$ & Measurement & Resistance & Predicted & Measured\\
\hline
\multirow{3}{*}
{$-$1} & \multirow{3}{*}
{1} & $R_{\textrm{long}}$ & $R_{14,23}$ & 0 & 0.039\\
 & & $R_{\textrm{Hall}}$ & $R_{13,24}$ & 0.5 & 0.666\\
 & & $R_{\textrm{mHall}}$ & $R_{42,31}$ & $-$0.5 & $-$0.630\\
\hline
\multirow{3}{*}
{1} & \multirow{3}{*}
{$-$1} & $R_{\textrm{long}}$ & $R_{14,23}$ & 0 & 0.013\\
 & & $R_{\textrm{Hall}}$ & $R_{13,24}$ & $-$0.5 & $-$0.522\\
 & & $R_{\textrm{mHall}}$ & $R_{42,31}$ & 0.5 & 0.583\\
\end{tabular}
\caption{Four-terminal resistances in Device B, measured at zero field, gate voltage $-8$ V, and 26 mK. Resistance values are given in units of $h/e^2$. Predicted values, given by the Landauer-B\"uttiker formalism, assume two chiral modes co-propagate along the domain wall.}
\label{tableb}
\end{table}

Dissipation along the domain wall appears thermally activated above $T=150$ mK (Fig.~2(b)). The longitudinal resistance $R_{14,23}$ increases with rising temperature according to an Arrhenius law $R_{14,23}\propto \exp (-T_0/T)$ with an activation gap $T_0=0.41$ K. Arrhenius activation with a comparable gap size is observed in the material's bulk conductivity $\sigma$, measured in the Corbino geometry~\cite{supinfo}. The temperature dependence of $R_{14,23}$ flattens below $T=150$ mK; the cause of this behavior is unclear. The consistent Arrhenius activation between $R_{14,23}$ and $\sigma$ suggests that bulk conduction causes dissipation in transport along the domain wall.

The effective Hall resistances $R_{13,24}$ and $R_{42,31}$ are oppositely-signed; further, their sign switches when the device's magnetic configuration is reversed. This confirms the chirality of conduction along the domain wall. The magnitudes of $R_{13,24}$ and $R_{42,31}$ at the optimum gate voltage, detailed in Table~II, are close to (but slightly exceed) $0.5h/e^2$, the expected value for two chiral modes. As shown in Fig.~3, $R_{13,24}$ and $R_{42,31}$ converge as the temperature rises, with $R_{13,24} \approx R_{42,31}\approx 0.25 h/e^2$ at 770 mK. Here, bulk conduction dominates and the two resistances are no longer analogous to Hall measurements; instead, they saturate at a positive value reflecting the sheet conductivity $\sigma$ and the device geometry.

\begin{figure*}
\centering
\includegraphics[width=0.85\textwidth]{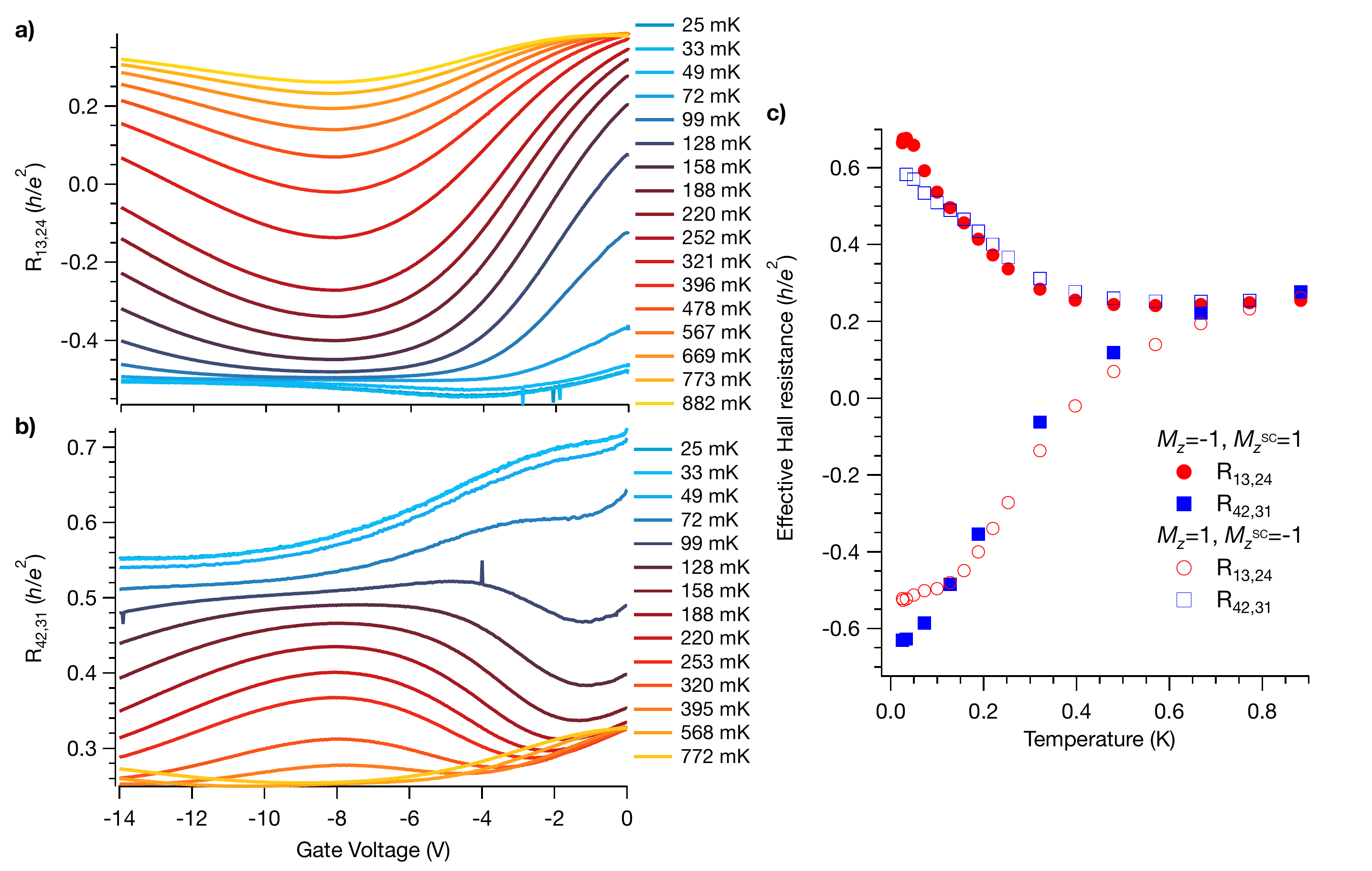}
\label{fig3}
\caption{Effective Hall resistances of Device B, a) $R_{13,24}$ and b) $R_{42,31}$, as a function of gate voltage at temperatures between 25 mK and 880 mK. Magnetization was $M_z=1$ and $M_z^{\textrm{SC}}=-1$, such that a domain wall with clockwise chirality connects the contacts. c) $R_{13,24}$ and $R_{42,31}$ at gate voltage $-8$ V, as a function of temperature, for the magnetic configurations $M_z=-1$ and $M_z^{\textrm{SC}}=+1$  (solid markers), and $M_z=1$ and $M_z^{\textrm{SC}}=-1$ (hollow markers).}
\end{figure*}

We have shown that magnetic domain walls in quantum anomalous Hall insulators conduct through chiral modes, which are expected to be topological in origin. Two modes are predicted to propagate along domain walls, thus we expect the effective Hall resistances of Device B to saturate at $h/2e^2$ at low temperatures. Though we did not observe such saturation, the measured effective Hall resistances at the optimum gate voltage are near the predicted value.

To explain the discrepancy, we propose that the magnetic exchange interaction, needed to open a gap in the MTI's surface states~\cite{Yu2010}, is reduced around the domain wall, allowing the formation of a compressible stripe. The stripe could perturb $\abs{R_{13,24}}$ and $\abs{R_{42,31}}$ from $h/2e^2$ without imparting a significant contribution to the longitudinal resistance~\cite{supinfo}. The width of the compressible stripe is presumably related to the spatial profile of the applied magnetic field, which should vary over hundreds of microns near the edge of the superconductor.

The authors acknowledge Andrew J. Bestwick for his contributions to the instrumentation and procedures used in this work, and Marc Kastner, Francesco Giazotto, Malcolm Beasley, and Inti Sodemann for insightful discussions. Device fabrication and measurement was supported by the U.S. Department of Energy, Office of Science, Basic Energy Sciences, Materials Sciences and Engineering Division, under Contract DE-AC02-76SF00515. Infrastructure and cryostat support were funded in part by the Gordon and Betty Moore Foundation through Grant No. GBMF3429. X. K., L. P. and K. L. W. acknowledge support from FAME, one of six centers of STARnet, a Semiconductor Research Corporation program sponsored by MARCO and DARPA; and from the Army Research Office under Grant Numbers  W911NF-16-1-0472  and W911NF-15-1-0561:P00001. X. K. acknowledges support from the Chinese National Thousand Young Talents Program and Shanghai Sailing Program under Grant Number 17YF1429200. Part of this work was performed at the Stanford Nano Shared Facilities (SNSF), supported by the National Science Foundation under award ECCS-1542152.

\clearpage

\pagebreak
\widetext
\begin{center}
\textbf{Chiral transport along magnetic domain walls in the quantum anomalous Hall effect}
\end{center}
\setcounter{equation}{0}
\setcounter{figure}{0}
\setcounter{table}{0}
\setcounter{page}{1}
\makeatletter
\renewcommand{\theequation}{S\arabic{equation}}
\renewcommand{\thefigure}{S\arabic{figure}}
\renewcommand{\bibnumfmt}[1]{[S#1]}
\renewcommand{\citenumfont}[1]{S#1}

\section{Supplemental Materials}

\subsection{Material growth}
Measurements were performed on high quality single crystalline (Cr$_{0.12}$Bi$_{0.26}$Sb$_{0.62}$)$_2$Te$_3$ films 6 quintuple layers (QLs) in thickness. Films were grown on a semi-insulating GaAs (111)B substrate in a Perkin-Elmer ultra-high-vacuum molecular beam epitaxy system. The substrate was annealed at $580^\circ$C in a Se-rich environment to remove the native oxide. The MTI film was grown with the substrate held at $200^\circ$C with the Cr, Bi, Sb, and Te source shutters simultaneously open. Growth was monitored using \emph{in situ} reflection high-energy electron diffraction (RHEED). After growth, a 2 nm Al layer was evaporated \emph{in situ} at room temperature and later allowed to oxidize in air to passivate the surface, protecting the film from unwanted environmental doping or aging effects.

\subsection{Device fabrication}
Devices were patterned using contact photolithography. For each patterning step, a hexamethyldisilazane adhesion layer was spin coated, followed by Megaposit SPR 3612 photoresist. The pre-exposure bake of $80^\circ$C for 120s was chosen to avoid thermal damage to the film. The photoresist was exposed under an ultraviolet mercury vapor lamp at approximately $70\textrm{ mJ}/\textrm{cm}^2$ and was developed in Microposit developer CD-30 for 35s. The device mesas were defined after patterning by etching the surrounding film with an Ar ion mill. Ohmic contacts were made by first cleaning the area with a brief exposure to an \emph{in situ} Ar ion source, and then evaporating 5 nm Ti and 100 nm Au, followed by liftoff. To realize a robust top gate, a dielectric was grown uniformly across the film by evaporating a 1 nm Al seed layer, which was allowed to oxidize, and then depositing approximately 40 nm of alumina by atomic layer deposition. The top gate was then fabricated by evaporating 5 nm Ti and 85 nm Au, followed by liftoff. Excess alumina dielectric on the surrounding area was etched using Microposit developer CD-26 (tetramethylammonium hydroxide based, metal ion free). Metal was evaporated using a Kurt Lesker electron beam evaporator with an \emph{in situ} Ar ion source. Atomic layer deposition used trimethylaluminum precursor and water as the oxidizer in a nitrogen purged vacuum chamber.

Device A was a Hall bar 500 $\mu$m in width. The connections between the length of the Hall bar and the voltage terminals were 20 $\mu$m wide. The pairs of voltage terminals within a domain (such as terminals 2 and 3) were separated by 400 $\mu$m, measured from center to center of the voltage terminals, meaning $R_{xx}$ and $R_{xx}^{\textrm{SC}}$ were measured across 0.8 squares. Terminals 3 (9) and 4 (8) were separated by 800 $\mu$m, meaning $R_{xx}^{\textrm{top}}$ and $R_{xx}^{\textrm{bottom}}$ were measured across 1.6 squares. Device A was not gate-tunable due to a short to one of the contacts through the alumina dielectric.

Device B was a triangular region of MTI film with side length $2.6\ \textrm{mm}$. Four contacts were placed within the triangular region; the MTI film was removed underneath each contact during the mesa etching step. Each contact was approximately $750\ \mu\textrm{m}$ tall and $300\ \mu\textrm{m}$ wide. The contacts were centered at a radius of about $1\ \textrm{mm}$ from a common origin in the center of the triangle, and were angularly separated by $20^\circ$. The angular separation was chosen to reduce the overall size of the device. The Fermi level was tuned in Device B using a top gate; the optimum gate voltage of $-8$ V was similar to that of other devices made from the same growth of MTI film.

\subsection{Lock-in measurement}
Four-terminal resistances were measured using a typical lock-in measurement setup in a dilution refrigerator at a base temperature of 25 mK. Devices A and B were measured in separate cooldowns. All measurements were current biased with a 5 nA AC bias, which was applied to the device by sourcing 5 V RMS across a 1 G$\Omega$ resistor.  The current traveling through the device and out of the drain terminal was measured with an Ithaco 1211 current preamplifier with 200 $\Omega$ input impedance, set to $-10^7$ V/A gain. Differential voltages $V_{xx}$ and $V_{yx}$ were measured with NF Corporation LI-75A voltage preamplifiers and Stanford Research Systems SR830 digital lock-in amplifiers. The excitation frequency was between 1 and 10 Hz; at higher frequencies, large phase differences developed between the excitation and measurement.

The amplifier chain in the measurement setup requires calibration for precise measurement due to uncertainty in the amplifiers' gains. Approximately four months prior to collection of data in this work, the amplifiers were calibrated to the Hall resistance of a quantum anomalous Hall device. Comparison of the QAH device's Hall resistance to the resistance of a standard resistor, using a cryogenic current comparator, verified its precise quantization to $h/e^2$. The amplifier chain used in lock-in measurements gives the uncalibrated measurement for the Hall resistance $V_{yx}/I = 26.75\ k\Omega$, which is $3.6\%$ higher than $h/e^2=25.81\ k\Omega$. All resistances presented in this work were multiplied by $0.965$ to correct for this inaccuracy.

\subsection{Meissner screening of the external field}
\begin{figure}
\centering
\includegraphics[width=0.45\textwidth]{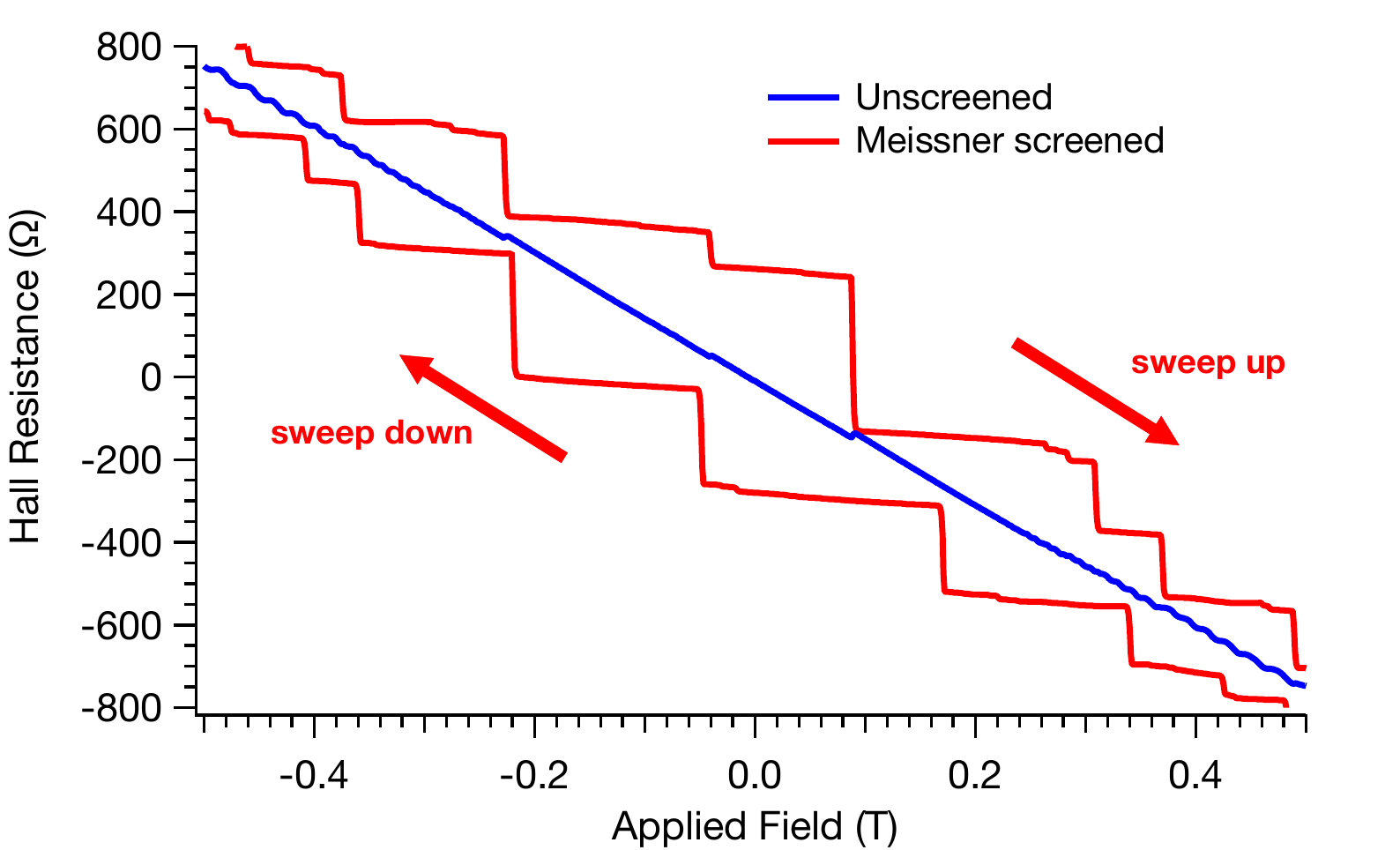}
\caption{Field sweep of the Hall resistance in a GaAs-based 2DEG, both away from (blue) and underneath (red) a superconducting cylinder placed on the surface of the device.}
\label{SupFig_GaAs}
\end{figure}

Reversal of the magnetization of the MTI film at dilution refrigeration temperatures occurs as the external field is swept over an approximately 50 mT window, between $H=150$ mT and $H=200$ mT. Therefore, to create a domain we must apply a 200 mT external field without the field underneath the superconducting cylinder exceeding 150 mT. To verify that niobium cylinders met this condition, a cylinder was placed on a Hall bar of a two-dimensional electron gas (2DEG) in a GaAs/Al$_{0.3}$Ga$_{0.7}$As heterostructure. Below the onset of Landau quantization, the Hall resistance away from the niobium cylinder was linear, with a Hall resistance of $1.53 \textrm{ k}\Omega/\textrm{T}$. Underneath the cylinder, the Hall resistance depended on the external magnetic field in a hysteretic stair-step pattern, as shown in Fig.~\ref{SupFig_GaAs}, likely reflecting vortex lattice pinning. Vortices depinned at a different series of external fields during every field sweep. Most steps persisted as the field was swept for at least 50 mT until the vortex lattice again depinned. The height of most steps corresponded to at least 50 mT of screened field (where the height is converted to units of field using the material's Hall slope). Therefore, the Meissner screening of the superconductor can screen 50 mT of field when the external field is 200 mT; accomplishing suitable screening, however, is dependent on how the vortex lattice depins on a given field sweep. The cylinder was made from unannealed commercial grade niobium of $99.5\%$ purity (Eagle Alloys Corporation). The 2DEG was 39 nm deep in a GaAs/Al$_{0.3}$Ga$_{0.7}$As heterostructure having two Si $\delta$-doping layers (dopant concentration $5\times 10^{12} \textrm{ cm}^{-2}$) at depths of 14 and 18 nm (IQE Corporation; $\mu=1.1\times 10^5\ \textrm{cm}^2/\textrm{Vs}$, $n=5.9\times 10^{11}\ \textrm{cm}^{-2}$).

\subsection{Field sweeps of screened devices}
\begin{figure}
\centering
\includegraphics[width=0.9\textwidth]{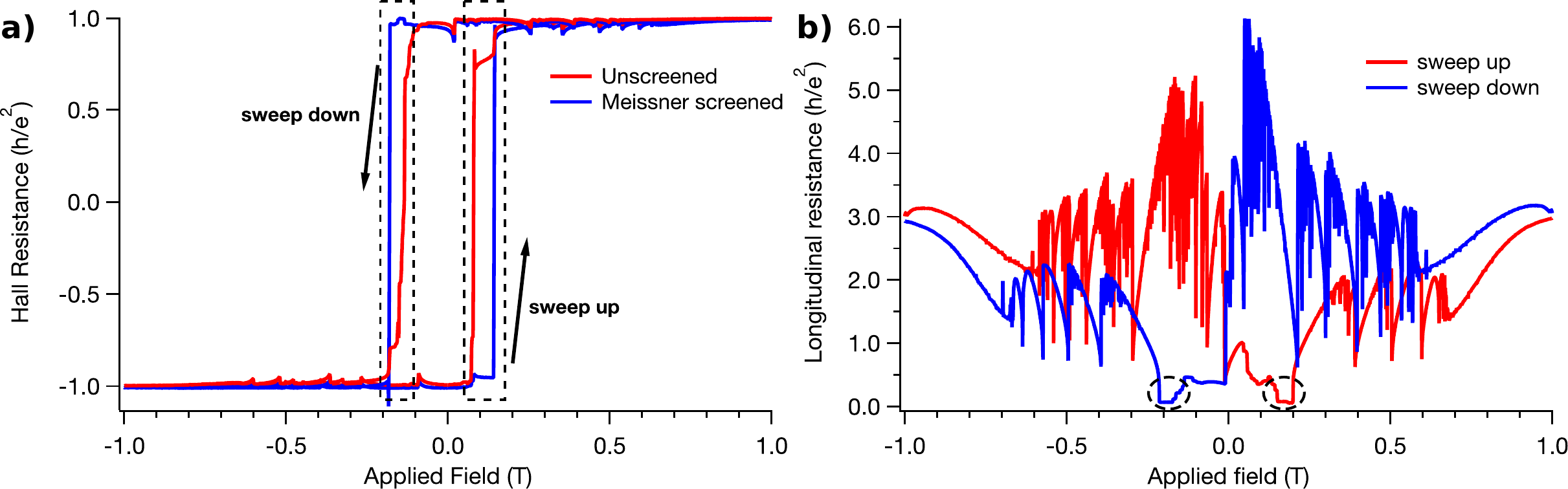}
\caption{a) Hysteresis curve of the Hall resistance in Device A, shown both away from (red) and underneath (blue) a superconducting cylinder placed on the surface of the device. b) Hysteresis curve of the longitudinal resistance in Device B. In both devices, the spikes in resistance are associated with vortex lattice depinning. Dashed areas indicate regions where a domain wall may be formed.}
\label{SupFig_field}
\end{figure}

The four-terminal resistances of Devices A and B are shown as the external magnetic field is swept in Fig.~\ref{SupFig_field}. The Hall resistance in the unscreened region of Device A changes sign at the coercive field, while the Hall resistance underneath the superconducting cylinder changes sign at a slightly higher field. A domain wall is created by sweeping the external field to a value within this window; the field is then brought back to zero, verifying the stability of the domain wall. In Device B, the minimum longitudinal resistance occurs within the same window, where conductive modes along the domain wall carry current between the four terminals. A domain wall was not formed on every field sweep, presumably because randomness in vortex lattice pinning caused the magnetization of the niobium cylinder to differ between field sweeps.

As the external field is swept, the measured voltages often change suddenly in value, and then slowly decay back to the original value. If the external field sweep is paused immediately after such a spike, the voltages nevertheless decay to their original values with the same time scale as when the sweep is not paused. We interpret these features as results of vortex lattice depinning in the niobium cylinder. As the external field is varied, the vortex lattice occasionally depins to reach an energetically favorable configuration for the new external field. When the vortex lattice depins, the superconductor releases heat into the MTI film. The longitudinal resistance in the MTI film is elevated until the device cools and reestablishes thermal equilibration with the bath. The transport of moderate-mobility GaAs 2DEGs in this range of field is less sensitive to temperature; therefore, lattice depinning had only a marginal effect when we measured a GaAs 2DEG screened by a superconducting cylinder.

\begin{figure}
\centering
\includegraphics[width=0.55\textwidth]{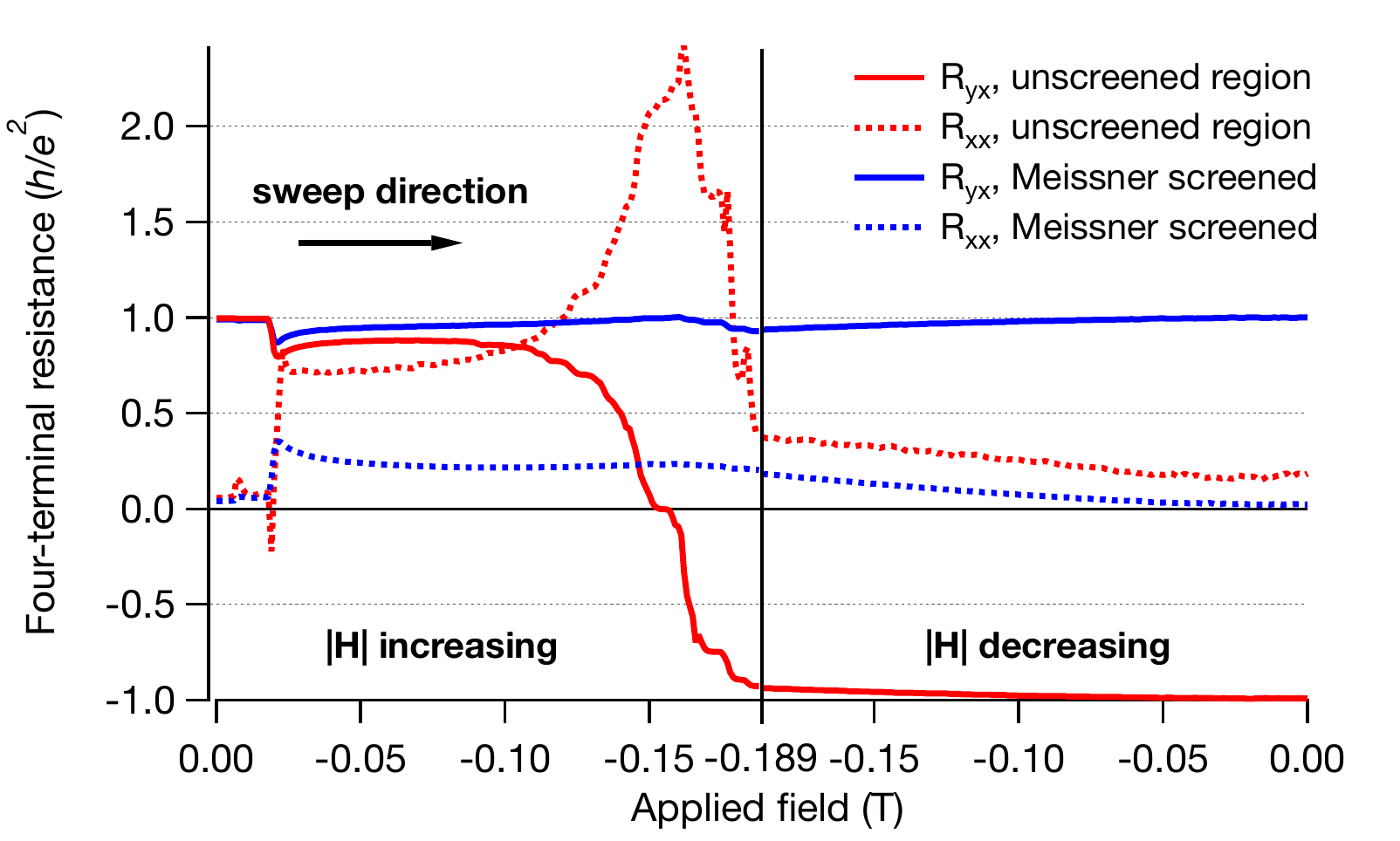}
\caption{Creation of a domain wall in Device A by sweeping the external field. An external field of $\mu_0 H=+1.2$ T was first applied to magnetize the entire device upwards $M_z=M_z^{\textrm{SC}}=+1$. The external field is swept to $\mu_0 H=-0.189$ T, switching the magnetization in the unscreened region $M_z=-1$, and is then swept back to zero. Hall (solid lines) and longitudinal (dashed lines) resistances are shown both underneath (blue lines) and away from (red lines) the superconducting cylinder.}
\label{SupFig_create}
\end{figure}

Fig.~\ref{SupFig_create} shows the process of creating a domain in Device A. The MTI begins having already been magnetized upwards by an external field $\mu_0 H=1.2$ T. The external field is swept to $-189$ mT, at which point the Hall resistance far from the niobium cylinder changes sign. The external field is then brought back to zero; the low longitudinal resistance inside each domain persists, demonstrating the stability of the magnetic domain wall.

\subsection{Predicted resistances in a Hall bar with magnetic domains}
The four-terminal resistances of a device may be computed using the Landauer-Buttiker formalism:
\begin{equation}
I_j =  \frac{e^2}{h}\sum_j \left(\bar{T}_{ij}V_j-\bar{T}_{ji}V_i \right)
\end{equation}
where the transmission coefficient $\bar{T}_{ij}=M_{ij}T_{ij}$ is the product of the number of modes $M_{ij}$ from terminal $j$ to terminal $i$, and their transmittance $0\leq T_{ij}\leq 1$. A uniformly magnetized Hall bar in the quantum anomalous Hall state has $M_{i-1,i}=1$ and $T_{i-1,i} = 1$ when the device's magnetization is upwards $M_z=1$, while $M_{i+1,i}=T_{i+1,i} = 1$ when $M_z=-1$; all other transmission coefficients are zero.

\begin{figure}
\centering
\includegraphics[width=0.35\textwidth]{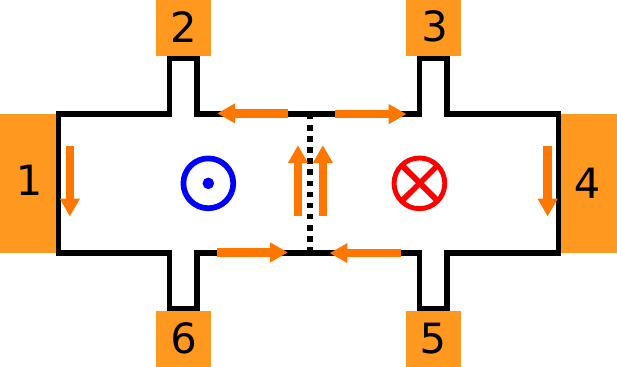}
\caption{A cartoon of a Hall bar with one magnetic domain wall, in the configuration $M_z=-1$, $M_z^{\textrm{SC}}=1$. The direction of magnetization is shown in red and blue, and the expected chiral modes are shown in orange. Since the chemical potential fully equilibrates between the two modes along the domain wall for $t=0.5$, the voltages measured at terminals 2 and 3 are equal.}
\label{SupFig_hbDomain}
\end{figure}

We calculate the longitudinal resistance across a domain wall as a function of the transmittance probability $t$ that a carrier entering a domain wall will exit the domain wall traveling in the in the other domain. Consider a carrier impinging on the domain wall illustrated in Fig.~\ref{SupFig_hbDomain} from the bottom left edge of the device. The carrier will travel along the domain wall to the top edge of the device, and either will leave the domain wall traveling rightwards, with transmittance probability $t$, on the opposite side of the domain wall, or will leave traveling leftwards, with transmittance probability $1-t$, staying within the original domain. For sufficiently long domain walls, we expect the modes to couple and fully equilibrate~\cite{upadhyaya2016}, such that a carrier leaving the domain wall has no memory of the side of the domain wall from which it entered; therefore, we expect $t=0.5$.

\begin{figure}
\centering
\includegraphics[width=0.45\textwidth]{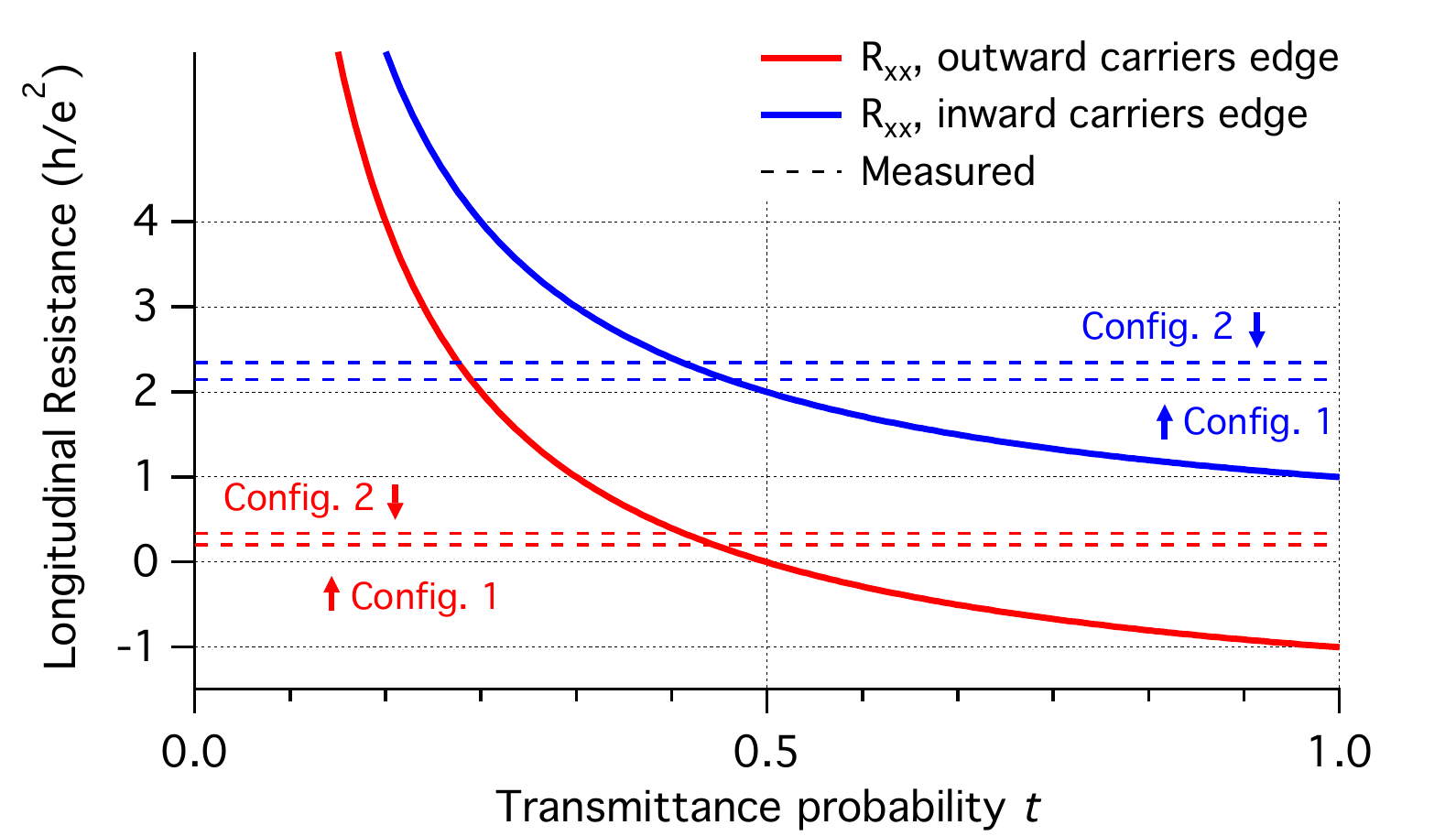}
\caption{Calculated longitudinal resistances (solid lines) across the domain wall $R_{xx}^{\textrm{top}}$ and $R_{xx}^{\textrm{bottom}}$, as a function of the transmittance probability $t$ across the domain wall, and measured longitudinal resistances (horizontal dashed lines), including measurements taken in both magnetic configurations. Edges are labeled by whether carriers propagate along the edge outwards from the domain wall, or inwards towards the domain wall. In the magnetic configuration $M_z=1$, $M_z^{\textrm{SC}}=-1$ (Config. 1), the edge with outward moving carriers is the bottom edge. When $M_z=-1$, $M_z^{\textrm{SC}}=1$ (Config. 2), the bottom edge has inward moving carriers.}
\label{SupFig_transmission}
\end{figure}

When $t=0.5$, carriers leaving a domain wall rightwards and leftwards have the same chemical potential, so the voltage along this edge of the device is zero (in Fig.~\ref{SupFig_hbDomain}, $V_3-V_2=0$). The computed resistances $R_{xx}^{\textrm{bottom}}=R_{14,65}$ and $R_{xx}^{\textrm{top}}=R_{14,23}$, assuming perfect chiral transport along all edges of the device, are shown as a function of transmittance probability $t$ in Fig.~\ref{SupFig_transmission}. For $t=0.5$, the Landauer-Buttiker formalism indeed gives $R_{xx}^{\textrm{bottom}}=2 h/e^2$ along the bottom edge of the Hall bar and $R_{xx}^{\textrm{top}}=0$ along the top edge of the Hall bar. The resistances switch when the magnetization of the device, and in turn the chirality of the domain wall, is reversed. The results for $t=0.5$ are close to the experimental results, as shown in Table~I of the main text, supporting that the modes co-propagating along the domain wall fully equilibrate. The two modes could equilibrate by intermixing either at the ends of the domain wall or along the length of the domain wall~\cite{upadhyaya2016}; our measurements do not distinguish between these two possibilities.

\subsection{Nonlocal resistances in a Hall bar with magnetic domains}
\begin{figure}
\centering
\includegraphics[width=0.55\textwidth]{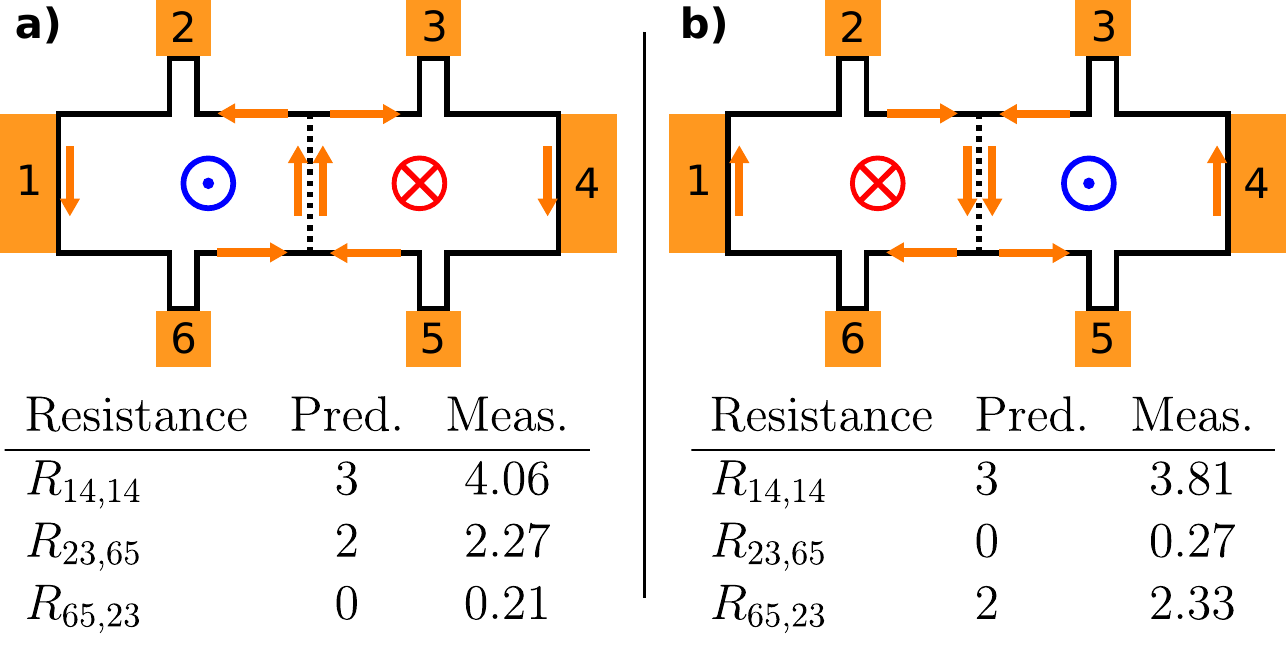}
\caption{Nonlocal and two-terminal resistances measured in Device A for the magnetic configurations a) $M_z=-1$, $M_z^{\textrm{SC}}=1$ and b) $M_z=1$, $M_z^{\textrm{SC}}=-1$. Four of the device's voltage terminals have been omitted from the schematic for clarity. The discrepancy between the measured and predicted values for the two-terminal resistance $R_{14,14}$ is partially due to the contact resistance of Ohmic contacts to the MTI.}
\label{SupFig_nonlocal}
\end{figure}

The Landauer-B\"uttiker formalism may further be used to calculate nonlocal resistances in Device A, where current is not sourced laterally across the Hall bar. The calculated and measured four-terminal resistances for a variety of nonlocal configurations are shown in Fig. ~\ref{SupFig_nonlocal}. Measurements were taken at 29 mK.

\subsection{Predicted resistances in Device B}

Four-terminal resistances in Device B may be predicted using the Landauer-B\"uttiker formalism. We consider a model where chiral modes co-propagate along the domain wall together with additional dissipative modes. We consider counterclockwise (as in the configuration $M_z=-1$, $M_z^{\textrm{SC}}=1$) chiral transport, and assume that there are $M_{i+1,i}=2$ modes each having transmittance $T_{i+1,i}=T_C$ from a terminal to the next. The transmission coefficients $\bar{T}_{ij}$ we may consider when adding dissipation to the model are constrained by the statement of conservation of current $\sum_i \bar{T}_{ij}=\sum_i \bar{T}_{ji}$.  To add dissipation, we add quasi-helical modes, whose conductance is $\bar{T}_{i\pm1,i}=\bar{T}_H$, between every adjacent terminal. The Landauer-B\"uttiker results for the effective longitudinal and Hall resistances along the domain wall, $R_L$ and $R_H$, are:

\begin{equation}
R_L=\frac{h}{e^2}\frac{\bar{T}_H}{2T_C^2+8T_C \bar{T}_H} \qquad R_H=\frac{h}{e^2}\frac{1}{2T_C+4 \bar{T}_H}+2R_L
\end{equation}

Comparing this result to the measured four-terminal resistances, shown in Table~II, suggests that the transmission from quasi-helical modes $\bar{T}_H$ is small compared to $e^2/h$, whereas the transmittance of each chiral mode is roughly $T_C\approx 0.75$. Perfect chiral conduction has transmittance $T_C=1$. We note that, while conservation of current requires that the chiral transmittance $T_C$ between each pair of terminals be equal, a device may have different quasi-helical conductances $\bar{T}^H_{i\pm1,i}$ for different terminals $i$.

We imagine a microscopic picture of transport in Device B to understand why $T_C\neq 1$ yet $\bar{T}_H$ is still small. Consider that the two chiral modes co-propagating along a domain wall are accompanied by a compressible stripe. Diffusive transport through the stripe adds a quasi-helical component to the device's transport $\bar{T}_H$. Since $\bar{T}_H$ is small, transport in the stripe must be highly diffusive, meaning the mean scattering length in the stripe is short compared to the distance between contacts. However, carriers may scatter between the chiral mode and the compressible stripe. Imagine a carrier leaving terminal $i$, traveling towards terminal $i+1$ in a chiral mode. Assuming that bulk conduction is negligible, the carrier must either reach terminal $i+1$ with probability $T_C$ or return to terminal $i$ with probability $1-T_C$. If the carrier scatters into the compressible stripe, it generally will not travel far before it scatters back into the chiral mode because the stripe is highly diffusive. This carrier will eventually reach terminal $i+1$. However, if the carrier scatters into the stripe soon after leaving terminal $i$, it may first return to terminal $i$ through diffusive transport in the stripe. Thus, this picture of a compressible stripe produces $T_C\neq 1$ while maintaining small quasi-helical transport. Macroscopically, this model preserves longitudinal resistance $R_L\approx 0$ along the domain wall, while producing non-quantized effective Hall resistances $\abs{R_H}>h/2e^2$, as observed in our measurements. 

\subsection{Four-terminal measurements in Device B}
\begin{figure}
\centering
\includegraphics[width=0.9\textwidth]{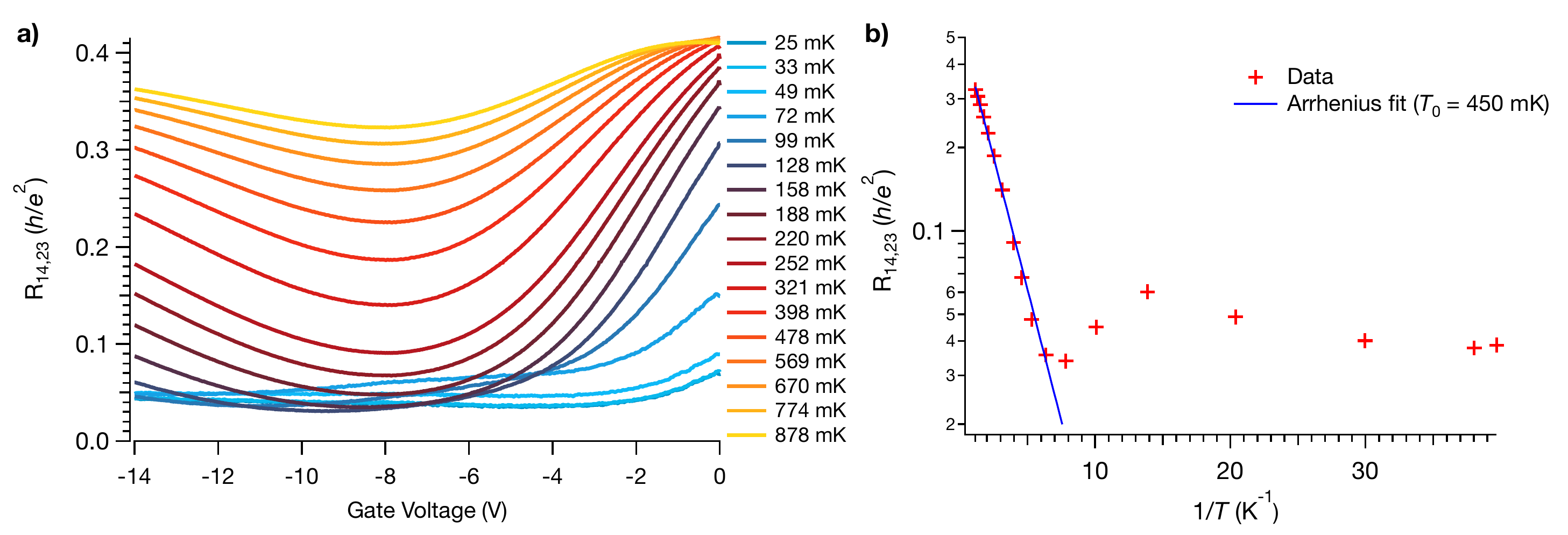}
\caption{a) Longitudinal resistance $R_{14,23}$ of Device B, as a function of gate voltage at temperatures between 25 mK and 880 mK. Magnetization was $M_z=-1$ and $M_z^{\textrm{SC}}=1$, such that a domain wall with counterclockwise chirality connects the contacts. b) Longitudinal resistance $R_{14,23}$ of Device B at gate voltage $-8$ V, shown on an Arrhenius plot. Only data at temperatures exceeding 150 mK are included in the fit to an Arrhenius law.}
\label{SupFig_domain7L}
\end{figure}

\begin{figure}
\centering
\includegraphics[width=0.9\textwidth]{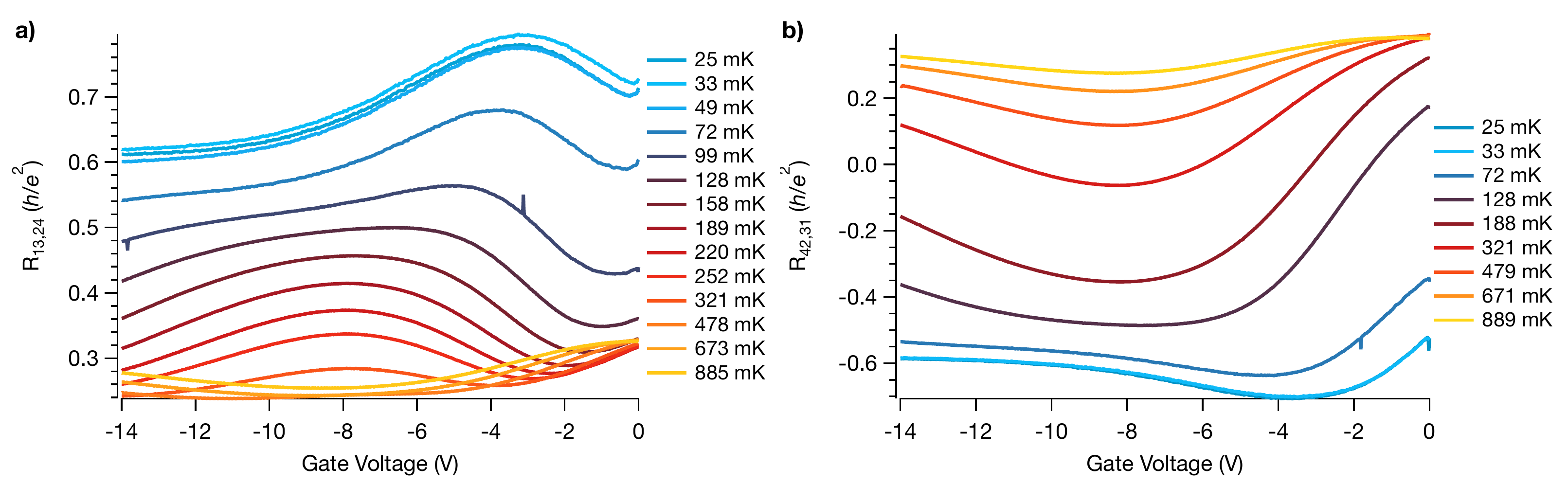}
\caption{Effective Hall resistances of Device B, a) $R_{13,24}$ and b) $R_{42,31}$, as a function of gate voltage at temperatures between 25 mK and 890 mK. Magnetization was $M_z=-1$ and $M_z^{\textrm{SC}}=1$, such that a domain wall with counterclockwise chirality connects the contacts.}
\label{SupFig_domain7H}
\end{figure}

The main text discussed four-terminal resistance measurements of Device B in the magnetization configuration $M_z=1$, $M_z^{\textrm{SC}}=-1$. The effective longitudinal resistance and effective Hall resistances were measured following separate external field sweeps, which created the magnetic domains. Four-terminal resistance measurements are shown for the magnetization configuration $M_z=-1$, $M_z^{\textrm{SC}}=1$ Fig.~\ref{SupFig_domain7L} and Fig.~\ref{SupFig_domain7H}. All measurements in this magnetic configuration followed the same field sweep.

\subsection{Arrhenius activation of bulk conduction}
\begin{figure}
\centering
\includegraphics[width=0.55\textwidth]{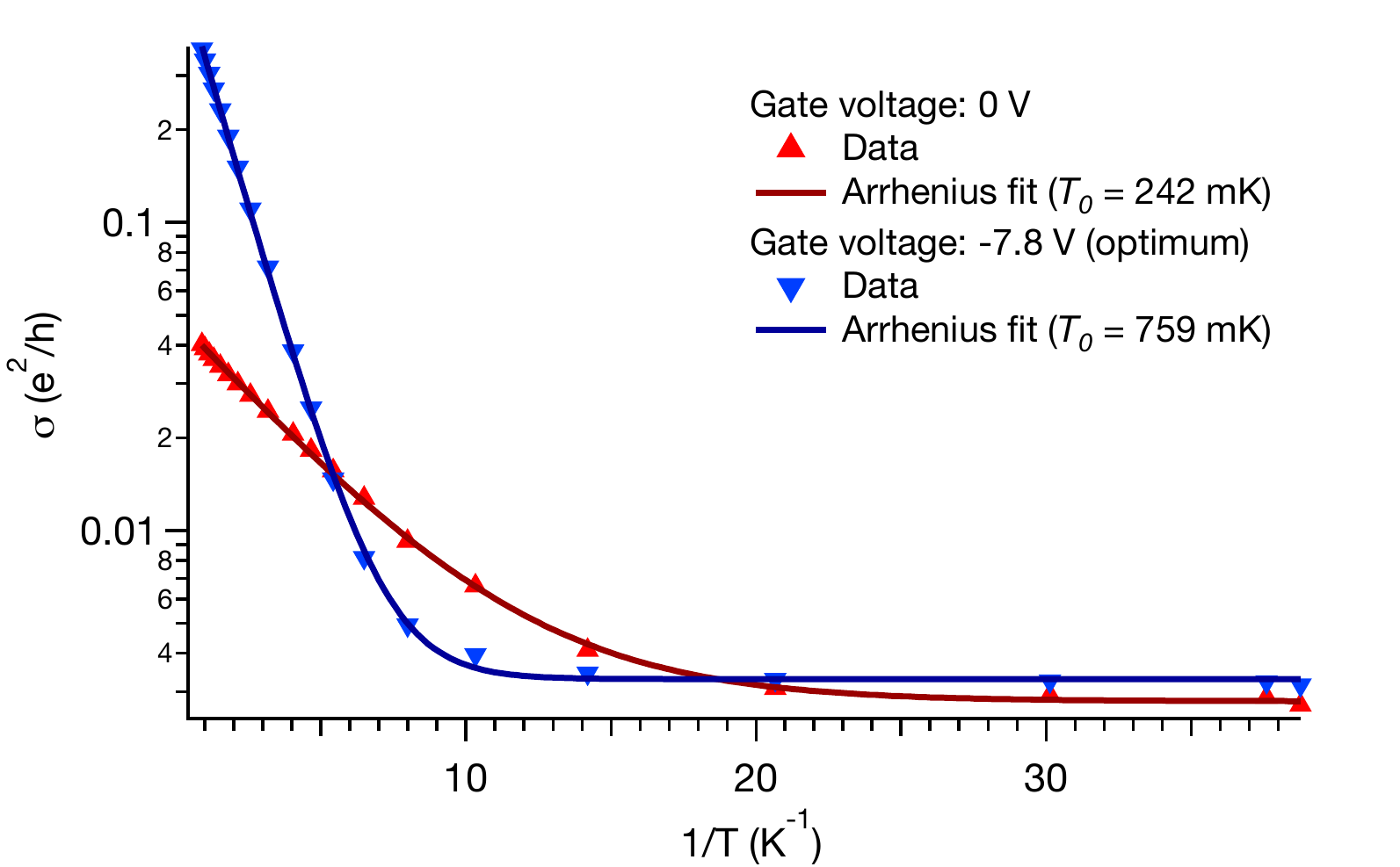}
\caption{The sheet conductivity of a Corbino geometry device, shown on an Arrhenius plot, at zero gate voltage (red) as well as the optimum gate voltage for the device, $-7.8$ V (blue). The device has an inner (outer) diameter of 200 $\mu$m (300 $\mu$m). The fits to an Arrhenius law include a constant offset term.}
\label{SupFig_corb}
\end{figure}

In the Corbino disk geometry, ohmic contact is made to the inner and outer rings of an annulus of a MTI film. No edge connects the two contacts, so the two-terminal conductivity of the device is a direct measurement of the bulk conductivity $\sigma$ (assuming low contact resistance). At low temperatures, the Corbino device is highly resistive when the Fermi level is tuned to the center of the gap by the top gate. As shown in Fig.~\ref{SupFig_corb}, the bulk conductivity increases with increasing temperature by an Arrhenius law with a constant offset $\sigma\sim e^{-T_0/ T}+c$. The activation gap, $T_0$, is largest when the gate is optimally tuned, at which point $T_0=759$ mK. At zero gate voltage, $T_0=242$ mK. The constant offset term $c$, which becomes dominant below 100 mK, may be caused by breakdown of the QAH effect under high electric fields.

\subsection{Jumps and plateaus in magnetic transitions}
\begin{figure}
\centering
\includegraphics[width=0.9\textwidth]{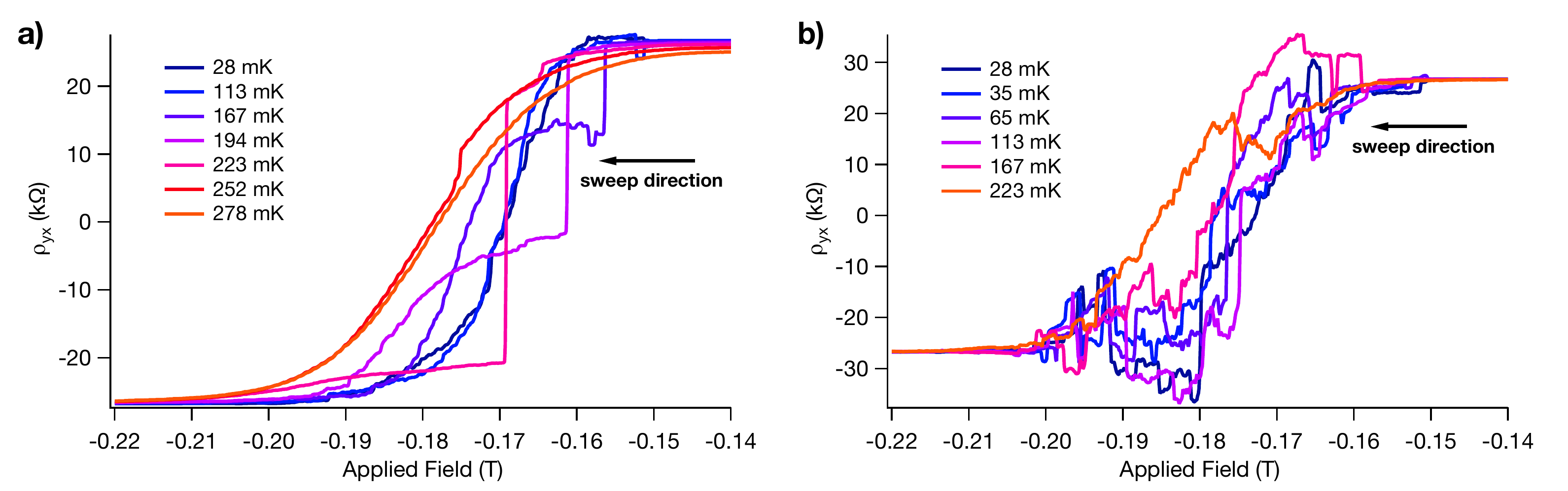}
\caption{Slow sweeps of the external field through the coercive field of Hall bars of a) $100\ \mu$m width and b) $20\ \mu$m width.}
\label{SupFig_jumps}
\end{figure}

The magnetization of an MTI film is reversed when the external field reaches the coercive field of the film, $H_C^{\textrm{TI}}$. Upon magnetic reversal, the film's Hall resistance changes sign. High resolution magnetic sweeps reveal that the Hall resistance does not smoothly transition, rather, it changes value in a series of jumps and plateaus. The jumps in $\rho_{yx}$ are largest at intermediate temperatures of around 220 mK. Above $T\approx 275$ mK, the Hall resistance varies smoothly. Such field sweeps are shown near the coercive field in Fig.~\ref{SupFig_jumps} for Hall bars having widths of $20\ \mu$m and $100\ \mu$m. Similar results have been reported recently in other (Cr$_{y}$Bi$_{x}$Sb$_{1-x-y}$)$_2$Te$_3$ films~\cite{Liu2016a}.

\subsection{Engineering magnetic domains}
We detail several alternative methods to realize magnetic domains in a QAH material at low temperatures. In this work, a large superconducting cylinder locally screens the external magnetic field. We used a niobium cylinder for its large critical field $H_{c1}^{\textrm{Nb}}$, the value of which reflects the maximum external flux that the cylinder can screen. Evaporated films of niobium may also have high critical fields, however, thin film superconductors do not effectively screen out-of-plane fields because of their large geometric demagnetization. In other words, for out-of-plane external fields, $H_{c1}^{\textrm{Nb}}=0$ in the limit of an infinitely thin superconducting film, and the film's magnetization is zero.

Whereas a superconductor locally screens the external field, a ferromagnet placed on a surface locally enhances the external field. To create a domain, the ferromagnet's stray field must add 50 mT (the width of the coercive transition) to the external field. To achieve this value, an iron or nickel film need be of order micron thickness. A ferromagnetic cylinder, machined from a bulk metal, potentially could create a magnetic domain. A related idea uses a thin film ferromagnet, magnetized in-plane, on the surface of an MTI film. The fringe field of the in-plane ferromagnet has a large out-of-plane component at its ends~\cite{Clinton1997,Castellana2006}, which potentially could create a magnetic domain in the MTI underneath the end of the ferromagnetic film. This method has the drawback that fringe fields are spatially narrow, so only a small domain could be formed.

Ideally, magnetic domain walls could be induced through the Oersted field from current passing through nanowires on the surface of an MTI. Such a device would feature transistor-like switching of current pathways through the MTI. Superconducting rather than normal metal nanowires are required to avoid Joule heating of the MTI. The critical fields of many superconductors far exceed the coercive field of the MTI, meaning enough current can pass through the superconductor to create a domain. The critical current of thin evaporated superconducting nanowires, however, is suppressed compared to the bulk material's critical current density. On the other hand, larger superconducting wires avoid suppressed critical current densities, but require high currents to generate the requisite Oersted field because of their larger radius.

\end{document}